%% file: main.tex
\documentclass[aps,prl,reprint,secnumarabic]{revtex4-2}

\usepackage{amsmath,amssymb}
\usepackage{graphicx}
\usepackage{microtype}
\usepackage{longtable}
\usepackage{needspace}
\usepackage[hidelinks]{hyperref}

\makeatletter
\let\frontmatter@footnote@produce\frontmatter@footnote@produce@footnote
\makeatother

\begin{document}

\raggedbottom

\title{QuOptLab: Advancing Quantum Optics Experiments with an AI Agent}
\author{Yuqing Li}
\email{yul658@pitt.edu}
\author{Zeguan Wu}
\email{zew79@pitt.edu}
\author{Yu Gan}
\email{yug130@pitt.edu}
\author{Min Chen}
\email{mic380@pitt.edu}
\author{Junyu Liu}
\email{junyuliu@pitt.edu}
\affiliation{University of Pittsburgh}

\begin{abstract}
Recent advances in AI-assisted theoretical research suggest that AI agents can contribute to research-level problems. Whether they can also advance experimental research by conducting experiments, analyzing the results, and proposing better experimental designs remains an open question. To address this question, we build a simulated quantum optics laboratory and integrate it with an AI agent in QuOptLab. The laboratory includes optical instruments, a robotic arm, and cameras, enabling the agent to carry out and iteratively refine experiments. We consider two complementary tasks: thin-film thickness measurement in free-space optics and edge-coupler design in integrated photonics. In the thin-film thickness measurement task, the agent repeatedly analyzes measurement results and revises its measurement method, including calibration and data analysis, reducing thickness-estimation error from 7.13 to 0.066~nm$^2$ over six epochs in a classical-interference run. For edge-coupler design, the agent iteratively proposes structural modifications, evaluates the resulting coupling performance, and uses the feedback to guide subsequent revisions, increasing simulated coupling efficiency from 85.81\% to 91.98\%.
\end{abstract}

\maketitle

\begin{figure*}[t]
    \centering
    \includegraphics[width=\textwidth]{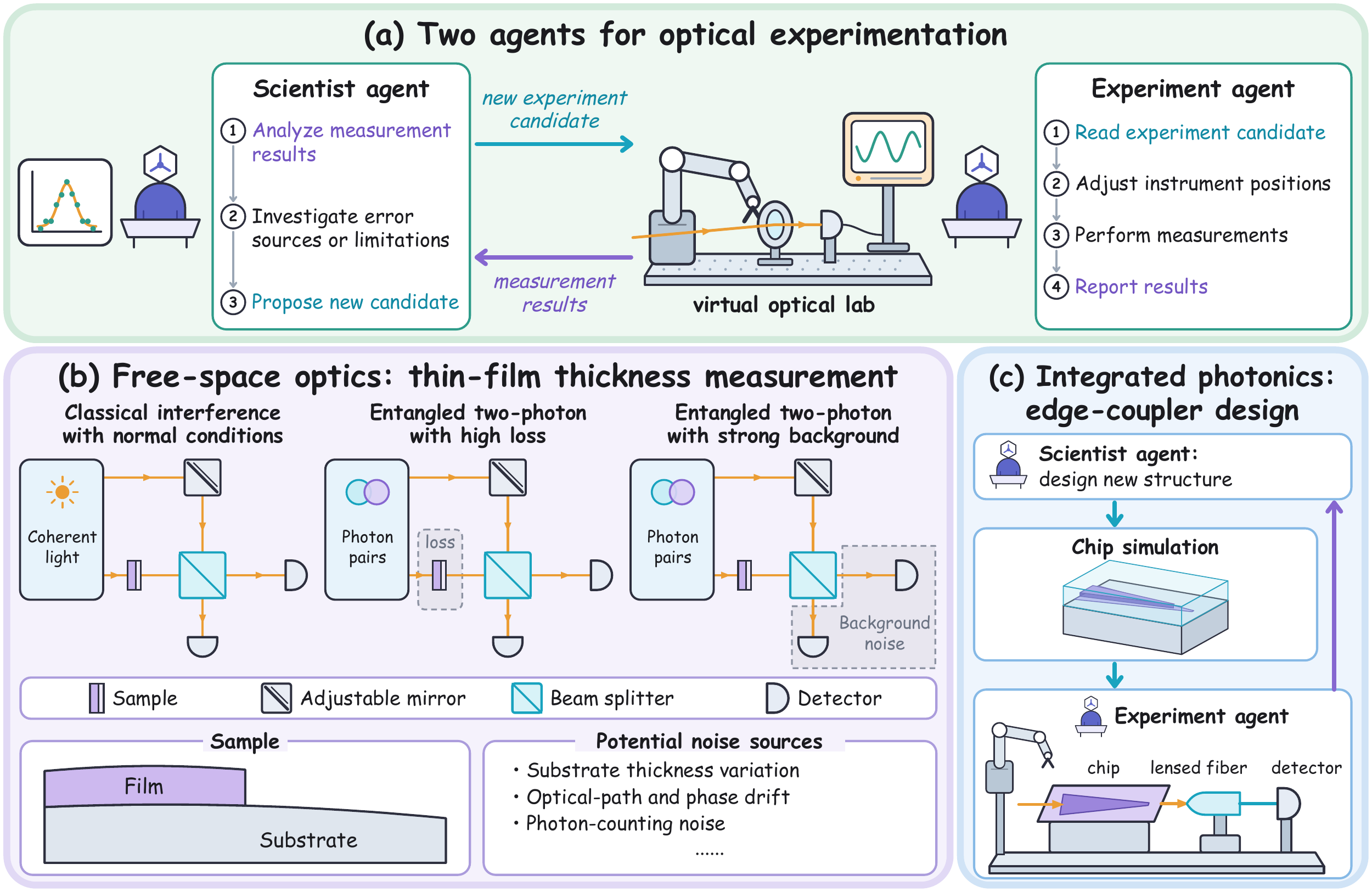}
    \caption{Overview of QuOptLab. (a) The scientist agent and experiment agent interact through a cycle of proposing new experiment candidates, performing measurements, and analyzing measurement results. (b) Free-space optics: thin-film thickness measurement. (c) Integrated photonics: edge-coupler design.}
    \label{fig:overview}
\end{figure*}

\section{Introduction}

AI has demonstrated the ability to contribute to theoretical research by addressing research-level problems through analysis and reasoning~\cite{Landreman2026Analytic,Alman2026Subquadratic,Issan2026Analytic,Li2026ClaimCentered}. In experimental research, AI agents have been used to carry out predefined experiments~\cite{Burger2020,Boiko2023,Uddin2025,Chen2026XrayScientist}. Our focus is on whether an AI agent can go beyond executing a given experiment and use measurement feedback to develop better measurement methods or device designs, thereby improving experimental performance. This requires the agent to analyze its observations, identify possible sources of error or limitations, propose changes to the experimental design, and assess their effects through further measurements. To investigate this question, we introduce QuOptLab, in which we build a virtual quantum optics laboratory for an AI agent equipped with robotic tools and cameras.

In our simulated laboratory, the agent is given an initial experimental setup and a performance objective. As illustrated in Fig.~\ref{fig:overview}(a), the agent operates in two complementary roles: a scientist agent and an experiment agent. The scientist agent analyzes measurement results, investigates error sources or limitations, and proposes new experiment candidates. The experiment agent uses information from cameras and optical measurements to guide instrument adjustments with the robotic arm, performs measurements, and reports the results to the scientist agent. The scientist agent uses these measurement results to revise the experiment candidate and guide the next epoch.

Our first experiment concerns thin-film thickness measurement in free-space optics (Fig.~\ref{fig:overview}(b))~\cite{Lyons2018HOM,Lualdi2025,Chen2019BeatNote}. A thin film changes the optical path of light passing through the sample, shifting the interference fringes. These shifts can be used to estimate the film thickness. We consider three subtasks: classical interference under normal conditions, entangled two-photon interference under high loss, and entangled two-photon interference under strong background. We generate simulated samples and divide them into a train set and a test set. The agent uses the known thicknesses of the train samples to analyze estimation errors and iteratively refine its measurement method. After each epoch, the revised measurement method is independently evaluated on the test set to assess whether it improves thickness estimation on unseen samples.

Our second experiment concerns edge-coupler design in integrated photonics (Fig.~\ref{fig:overview}(c)). Edge couplers transfer light between on-chip waveguides and optical fibers through the end facet. Their design aims to reduce optical loss by gradually transforming the tightly confined waveguide mode into a field that matches the fiber mode~\cite{Hu2021Coupler,Shi2025}. In this experiment, the simulated laboratory provides a robotic arm and optical measurement apparatus, while chip realization is handled by a backend electromagnetic simulator. After the agent proposes a modified chip design, the backend simulates its optical response and makes the resulting virtual chip available in the laboratory. The agent then uses the robotic arm and measurement apparatus to align the optical setup and measure the chip's coupling performance. These measurements provide feedback for analyzing the proposed design and deciding how to modify it in the next epoch.

In summary, we build a virtual optical laboratory equipped with a robotic arm to evaluate an AI agent's ability to improve experimental performance across two distinct optical experiments. We evaluate QuOptLab using GPT-6 Astra as the underlying model for the agent. In the thin-film thickness measurement experiment, we compare test-set mean squared error (MSE) between the first and sixth epochs. The runs with the largest relative reductions in each subtask reduce MSE from 7.13 to 0.066~nm$^2$ under normal classical-interference conditions, from 4.94 to 2.08~nm$^2$ under high loss, and from 3.43 to 0.421~nm$^2$ under strong background. In the edge-coupler design experiment, the best candidate increases simulated coupling efficiency from 85.81\% to 91.98\% and reduces spectral variation from 0.855 to 0.154~dB.

In this paper, Section~\ref{sec:related-work} reviews related work, and Section~\ref{sec:quoptlab} introduces QuOptLab. Sections~\ref{sec:thin-film} and~\ref{sec:edge-coupler} present the two experiments. Section~\ref{sec:conclusion-discussion} concludes the paper and discusses limitations and future directions.

\section{Related Work}
\label{sec:related-work}
Robotic laboratories have received considerable attention in experiment-intensive fields such as biology and chemistry, with applications including reaction optimization~\cite{Burger2020,Boiko2023}, hypothesis-driven studies of gene function~\cite{King2009}, and drug discovery~\cite{Williams2015}. In optics, Uddin et al.~\cite{Uddin2025} demonstrated robotic assembly, alignment, measurement, and feedback-based setup optimization in a real laboratory. For quantum experiments, k-agents~\cite{Cao2025} uses measurement feedback to guide superconducting-qubit calibration and operation. In concurrent work, QORA~\cite{Yan2026QORA} combines simulation and experiments to discover quantum-control sequences that improve coherence in diamond nitrogen-vacancy ensembles. Our work focuses on how an AI agent uses measurement results to improve quantum optics experiments.

\section{QuOptLab: Agent with Its Virtual Laboratory}
\label{sec:quoptlab}

QuOptLab is a virtual optical laboratory with an AI agent, enabling it to iteratively improve experimental performance toward a specified objective, starting from an initial experimental setup.  As illustrated in Fig.~\ref{fig:overview}(a), we organize this process into two complementary roles: a \textit{scientist agent} and an \textit{experiment agent}.

\paragraph{Scientist agent.} The scientist agent analyzes measurement results to identify possible sources of error or limitations and uses these findings to propose new experiment candidates. Each new experiment candidate specifies what to change, why the change may improve the experiment, and what measurements are needed to assess it.

\paragraph{Experiment agent.} The experiment agent reads the new experiment candidate, adjusts instrument positions, performs the specified measurements, and returns the measurement results and relevant experimental conditions to the scientist agent for analysis.

To be specific, the laboratory provides three fixed cameras---one overhead and two oblique---and a wrist-mounted camera. ArUco markers serve as visual labels on the magnetic base and height-adjustable head of each movable optical module~\cite{GarridoJurado2014Aruco}. A vision program detects these markers in camera images, combines observations from multiple views to estimate component positions, orientations, and uncertainties, and returns the results to the experiment agent. The agent decides what operation to perform, such as moving a particular module to a specified position or turning a knob by a specified number of revolutions. Before execution, the manipulation program verifies that the estimates remain valid and that the positional uncertainty is within the permitted threshold. If either check fails, the program rejects the operation and returns the reason to the agent, prompting it to obtain new observations. Otherwise, the program uses the estimated module poses and the known locations of handles and knobs on each module to determine where to grasp or make contact. It then computes joint motions through inverse kinematics and controls a simulated seven-degree-of-freedom xArm7 with a gripper in MuJoCo~\cite{Todorov2012MuJoCo,MenagerieXArm7}. The simulation monitors collisions, checks whether the gripper or tool actually reaches the target, and returns the execution results to the agent. Example views of the virtual laboratory are shown in Appendix~\ref{app:lab-views}.

The experiment agent uses two types of feedback to decide its next action. During apparatus assembly, it compares the position and orientation estimates returned by the vision program with the intended layout to check whether components have been placed as expected. Together with feedback on whether operations succeed or fail, these estimates guide its decisions to continue assembly, adjust component positions, or request new observations. During optical alignment, the agent compares photon counts or optical-power readings before and after knob adjustments to assess whether alignment has improved. It can continue in the same direction when readings increase, reverse or change direction when they decrease, and reduce the step size near a promising setting.

\section{Experiment I: Thin-Film Thickness Measurement}
\label{sec:thin-film}
\subsection{Experimental Setting}

Our first experiment studies whether an AI agent can improve a measurement method for estimating thin-film thickness in free-space optics. The virtual laboratory provides coherent-light and photon-pair sources, mirrors, beam splitters, dichroic mirrors, wave plates, fiber couplers, photon-counting detectors, power meters, beam-profiling cameras, adjustable irises, and beam dumps, together with sample translation stages, optical delay stages, and piezo-mounted mirrors. Sample translation, delay-stage motion, and piezo displacement are electronically controlled. For selected components, including mirrors and fiber couplers, both manual and motorized mounts are available. Motorized axes are adjusted through electronic control commands, whereas manual adjustments require the experiment agent to use the robotic arm to engage and turn the corresponding knobs. Wave-plate rotation and iris adjustment also require robotic manipulation. Table~\ref{tab:available-apparatus} in Appendix~\ref{app:apparatus} lists the available components and their control modes.

For this experiment, we generate simulated samples with film thicknesses drawn from 470--530~nm using stratified sampling and introduce random variations in the substrate slope to model sample-dependent substrate errors. As illustrated in Fig.~\ref{fig:overview}(b), each sample consists of a partially coated substrate, allowing the agent to measure both coated and uncoated regions. We consider three subtasks: classical interference under normal conditions, entangled two-photon interference under high loss, and entangled two-photon interference under strong background. Before any measurements, we select a fixed subset of five train samples with known ground-truth thicknesses and twenty test samples reserved for independent evaluation from the pre-generated sample pool. All three subtasks use these same train and test sets. We evaluate thickness-estimation accuracy using the MSE between estimated and ground-truth thicknesses across repeated measurements of the test samples.

\begin{figure*}[t]
    \centering
    \includegraphics[width=\textwidth,trim=0 15.53bp 0 0,clip]{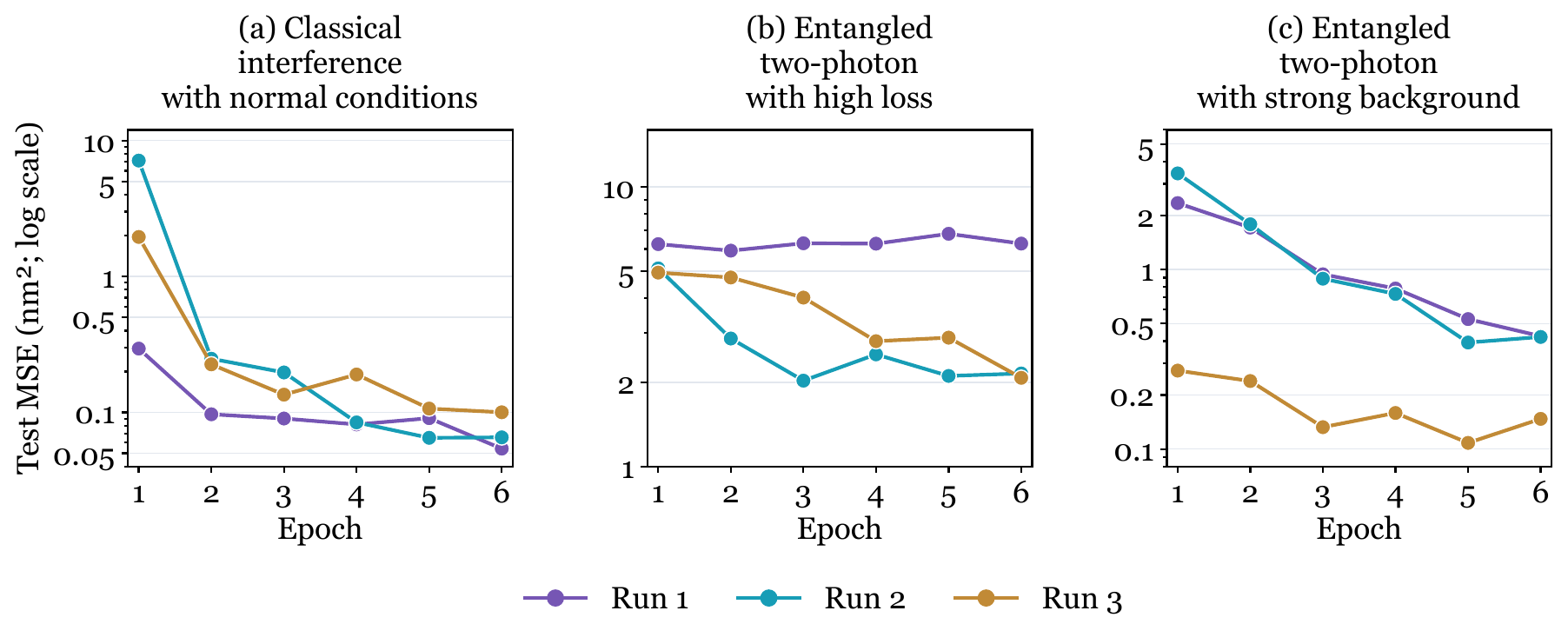}
    \caption{Test-set MSE across six epochs.}
    \label{fig:thickness-mse-rounds}
\end{figure*}

The first subtask uses classical interference under normal conditions. A piezo-mounted mirror is displaced to vary the optical path length in one arm of the interferometer, thereby scanning the relative phase between the two arms. The photon counts recorded at the two output ports oscillate approximately sinusoidally with mirror displacement; fitting these oscillations yields the fringe phase, whose change between the coated and uncoated sample regions is used to estimate film thickness. Substrate variation, optical-path drift, and photon-counting noise can complicate the extraction of the film contribution.

The second and third subtasks use energy-entangled two-photon interference under high-loss and strong-background conditions, respectively, with an initial setup adapted from Lualdi et al.~\cite{Lualdi2025}. The source generates pairs of photons with different wavelengths, approximately 810 and 1550~nm. Each pair enters the interferometer in a superposition of two possibilities: the shorter-wavelength photon travels through one arm and the longer-wavelength photon through the other, or vice versa. After the two paths are recombined, wavelength-selective optics at each output port separate the two photons by wavelength for detection. This gives two detectors per output port and four detection channels in total. Rather than measuring only individual photon counts, we count detection events that occur in pairs within a short time window, referred to as coincidence counts. As in the classical-interference experiment, a piezo-mounted mirror is displaced to vary the optical path length in one arm of the interferometer. The coincidence counts oscillate with mirror displacement, producing two-photon interference fringes that carry information about film thickness. Two-photon interference can retain high fringe visibility under unequal path loss, while coincidence detection suppresses uncorrelated background. The measurement principle and its robustness against unequal path loss and uncorrelated background are explained in Appendix~\ref{app:two-photon-principle}.

\subsection{Experimental Results}

In classical interference under normal conditions (Fig.~\ref{fig:thickness-mse-rounds}(a)), the agent improves thickness estimation by revising both the measured observables and the data analysis. In run 2, it finds that comparing fringe phases across uncoated and coated regions produces unstable estimates. It therefore shifts the main thickness-sensitive measurement to the fringe-amplitude profile near the film edge, while retaining measurements in both regions as references for amplitude normalization. Together with revised sampling and calibration, this change reduces test-set MSE from 7.13 to 0.248~nm$^2$ between the first and second epochs. Subsequent revisions concentrate more measurement time near the edge and account for shared piezo-position errors. The agent then finds that residual sample-dependent errors correlate with spatial phase gradients and hypothesizes that substrate variation contributes to these errors. It adds a partial phase-gradient correction and updates calibration, reaching an MSE of 0.066~nm$^2$ in the sixth epoch. In addition, run 1 combines edge-amplitude estimation with a measured-phase correction, while run 3 jointly models the film response, spatial phase variation, and temporal drift. Appendix~\ref{app:agent-classical} summarizes the changes made in each epoch.

\begin{figure*}[t]
    \centering
    \includegraphics[width=\textwidth]{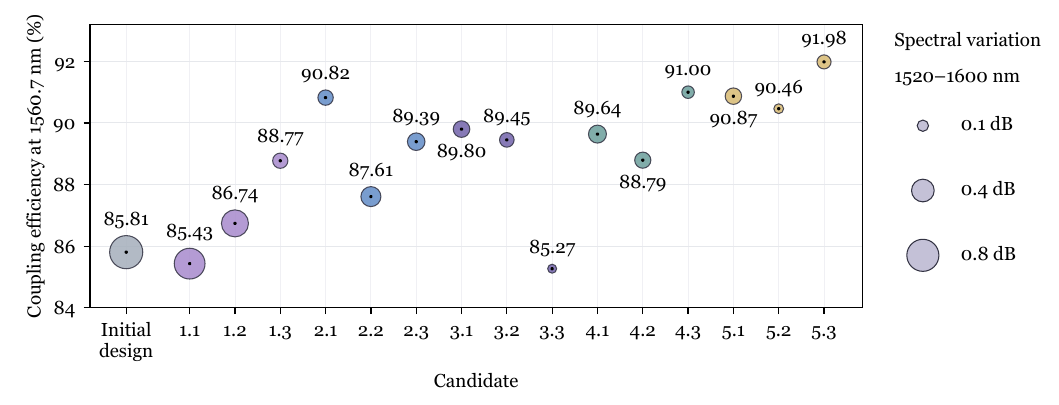}
    \caption{Simulated coupling efficiency across five design epochs. Smaller circles indicate lower spectral variation and greater spectral stability.}
    \label{fig:coupler-efficiency-stability}
\end{figure*}

In entangled two-photon interference under high loss (Fig.~\ref{fig:thickness-mse-rounds}(b)), the agent analyzes repeated measurements of the train samples and identifies limited photon counts as an important constraint on measurement precision. It improves optical alignment to collect more photon pairs and revises the sampling method to extract more phase information within the available measurement time. Run 2 concentrates measurements at sensitive phase working points and alternates between nearby coated and uncoated regions, while accounting for temporal drift. Run 3 develops a matched two-phase sampling method that reduces phase-fitting uncertainty. Run 1 improves photon collection but retains a substrate-dependent bias identified in our subsequent analysis, illustrating that increasing counts alone does not resolve all sources of error. The sampling and analysis methods are described in Appendix~\ref{app:agent-loss}.

In entangled two-photon interference under strong background (Fig.~\ref{fig:thickness-mse-rounds}(c)), the agent analyzes repeated measurements to investigate fluctuations in the thickness estimates. In run 3, it observes that the fringe amplitude and background-corrected mean coincidence level fluctuate together, while their ratio remains relatively stable. This suggests that variations in signal intensity contribute to the estimation error. The agent therefore uses normalization to reduce sensitivity to intensity fluctuations, while subtracting estimated accidental coincidences to suppress background contributions. It then combines thickness estimates derived from the normalized fringe amplitude and mean coincidence level. Details of the measurement method are provided in Appendix~\ref{app:agent-background}. Whereas the high-loss subtask emphasizes photon collection and efficient phase sampling, this method emphasizes background correction and normalization. Together, these subtasks illustrate how the agent uses measurement feedback to adapt its measurement method to different noise conditions.

\begin{figure*}[!t]
    \centering
    \includegraphics[width=\textwidth,trim=0 24.63bp 0 0,clip]{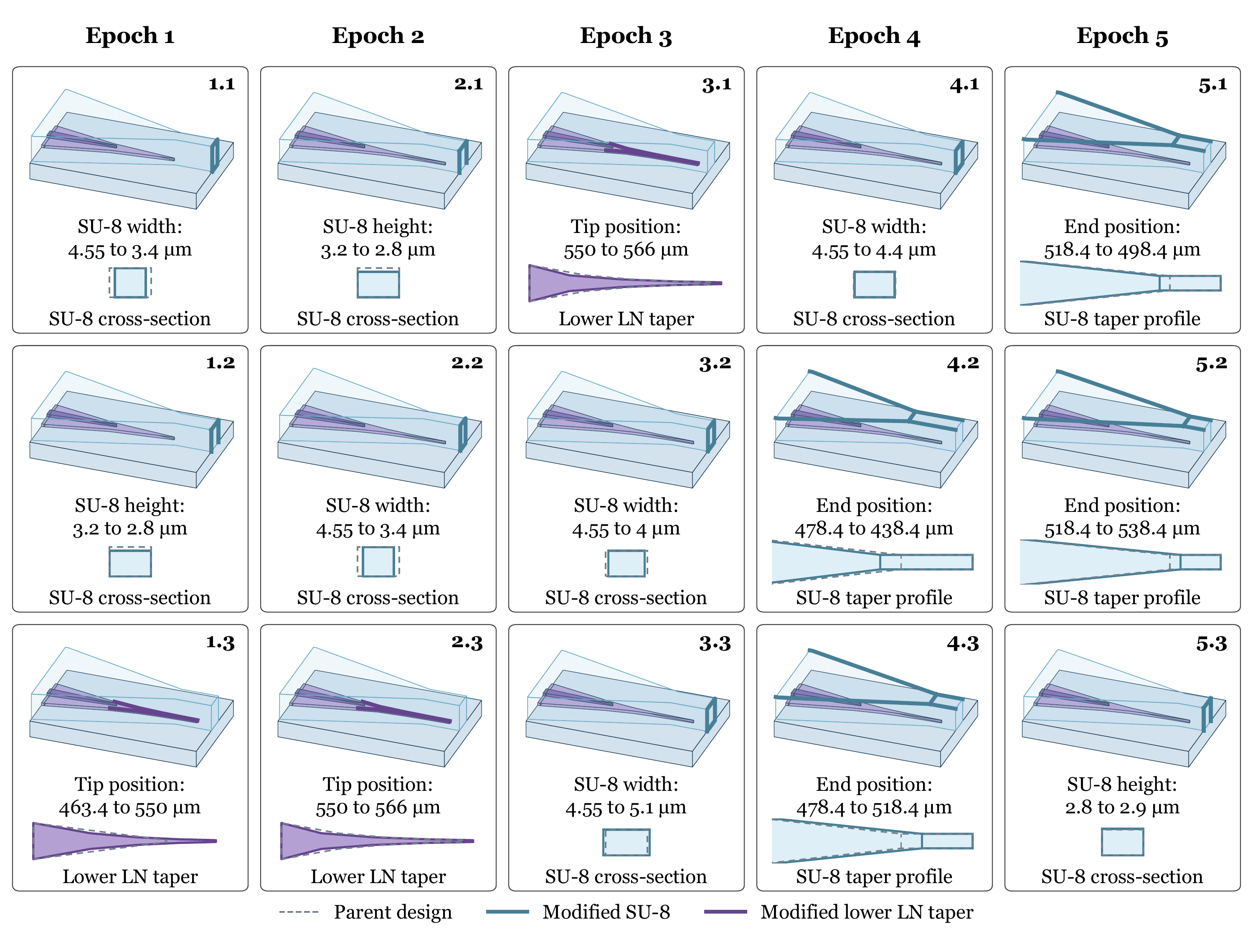}
    \caption{Fifteen candidate structures across five design epochs.}
    \label{fig:coupler-nine-structures}
\end{figure*}

\section{Experiment II: Edge-Coupler Design}
\label{sec:edge-coupler}

\subsection{Experimental Setting}

Our second experiment studies whether an AI agent can improve chip-to-fiber coupling by revising an edge-coupler design based on simulated measurement results. The initial design is adapted from Shi et al.~\cite{Shi2025} and consists of two stacked lithium niobate (LN) tapers covered by a waveguide made from SU-8, an epoxy-based negative photoresist, on an oxide-on-silicon substrate. The agent can modify the geometry and structure of the coupling region, including the shapes of the LN tapers and the dimensions of the SU-8 waveguide. The materials and LN tip widths remain fixed. The tip-width constraints retain the minimum feature sizes specified by the fabrication design rules, while allowing changes to taper profiles and lengths. The objective of the agent is to increase coupling efficiency at 1560.7~nm and improve spectral stability over 1520--1600~nm. Spectral variation is the difference between the maximum and minimum coupling efficiencies in dB over 1520--1600~nm.

Candidate chips are simulated using three-dimensional finite-difference time-domain (FDTD) calculations in Ansys Lumerical. The mesh is refined in the waveguide region and its immediate surroundings, with coarser transverse spacing farther away to reduce computational cost. Across five design epochs, the agent proposes three candidates per epoch and receives the simulation and evaluation results for all three to guide subsequent revisions.

\subsection{Experimental Results}

Chip-to-fiber coupling depends on both the transmission of light through the on-chip waveguide and taper, and the matching of the output field to the receiving fiber mode. Loss can therefore arise before light reaches the end facet or when the emitted field is not efficiently coupled into the fiber. These two considerations underpin established edge-coupler designs that combine gradual mode conversion with output-mode shaping~\cite{Hu2021Coupler,Shi2025}.

Figure~\ref{fig:coupler-nine-structures} summarizes the candidate geometries and their structural modifications, with each candidate labeled by its epoch and its index within that epoch. After the first epoch, the agent observes that reshaping the lower LN taper in candidate 1.3 increases the power coupled into the receiving fiber mode, while changing the total end-facet power only slightly. It therefore hypothesizes that improved mode matching between the output field and the receiving fiber mode may account for much of the gain. Guided by this hypothesis, the agent tests changes to the SU-8 height and end-facet width, together with further adjustments to the lower LN taper profile, during the second and third epochs.

In the fourth epoch, the agent retains the previously revised lower LN taper and reduced SU-8 height, further adjusts the end-facet width in candidate 4.1, and tests shorter and longer SU-8 taper sections in candidates 4.2 and 4.3 while keeping the end-facet dimensions unchanged. Candidate 4.2 produces more total end-facet power than candidate 4.3 but couples less power into the receiving fiber mode, further supporting the mode-matching hypothesis. In the fifth epoch, the agent tests nearby taper lengths in candidates 5.1 and 5.2 and increases the SU-8 height from 2.8 to 2.9~$\mu$m in candidate 5.3. As shown in Fig.~\ref{fig:coupler-efficiency-stability}, candidate 5.3 yields the highest simulated coupling efficiency among the fifteen candidates, reaching 91.98\%, compared with 85.81\% for the initial design. Fig.~\ref{fig:coupler-efficiency-stability} also shows reduced spectral variation across all fifteen candidates relative to the initial design, indicating a flatter coupling response over 1520--1600~nm.

Following the five evaluated epochs, we ask the agent to further analyze the remaining loss. For candidate 5.3, the reported fiber-coupled power is approximately 97.9\% of the total power just outside the end facet, suggesting that relatively little of the transmitted power remains unmatched to the receiving fiber mode. The agent therefore turns its attention to the remaining transmission loss. It observes similar internal net-power readings across several geometrically different candidates and, together with an approximate Fresnel-reflection estimate, hypothesizes that reflection at the SU-8--air end facet may be a major contributor. However, our simulated experiment uses a fixed material stack and does not permit adding an end-facet coating. Extensions to the simulation framework for evaluating such modifications are discussed in Section~VI.2.

\section{Conclusion and Discussion}
\label{sec:conclusion-discussion}

\subsection{Conclusion}
We built a virtual quantum optics laboratory and integrated it with an AI agent in QuOptLab to study how the agent can advance experiments through measurement feedback. The agent revises measurement methods by analyzing estimation errors, with the largest relative improvements across the three thin-film subtasks reducing test-set MSE from 7.13 to 0.066, 4.94 to 2.08, and 3.43 to 0.421~nm$^2$ between the first and sixth epochs. The high-loss and strong-background subtasks further illustrate the agent's ability to identify likely sources of measurement error, analyze their effects, and adapt its measurement method to different noise conditions. In edge-coupler design, its mode-matching hypothesis guides changes to the taper and SU-8 geometry, increasing simulated coupling efficiency from 85.81\% to 91.98\% in the best candidate. These results illustrate how the agent uses experimental observations to guide improvements in measurement methods and device designs.

\subsection{Discussion}

In this section, we discuss the limitations of our study and directions for future work.

First, although the agent can revise the geometry and structure of the coupling region in the second experiment, its proposals across the five evaluated epochs remain within the initial two-level taper architecture, modifying taper profiles and SU-8 dimensions rather than introducing a substantially different structure~\cite{Liu2025Bident,Liang2023Trident,Chen2024ThreeDimensional}. The agent proposes three candidates per epoch, and its output indicates that it is cautious about major structural changes and favors smaller modifications whose effects are easier to interpret. We also keep the materials fixed, since allowing material substitution would add substantial modeling and validation complexity to the current simulation pipeline. The end-facet-reflection hypothesis further motivates extending the simulation framework to evaluate antireflection coatings. Future work will provide more design epochs and larger simulation budgets, and extend the material models, to explore broader structural and material changes.

Second, both experiments are conducted in simulation, so the observed improvements remain dependent on the assumptions of the virtual laboratory and its optical models. Their effectiveness in a physical laboratory has yet to be established. For the integrated-photonics task, we plan to fabricate and experimentally characterize promising agent-designed couplers to assess whether the simulated gains persist under fabrication variations and real measurement conditions.

Finally, the physical consistency of the virtual laboratory requires further systematic verification. Future work will combine automated checks of the simulation programs with formal verification of selected components against their governing equations and stated assumptions. These efforts will complement numerical convergence studies and experimental validation to strengthen confidence in the physical fidelity of the virtual laboratory.

\section*{Acknowledgments}
YL, ZW, YG, MC, and JL are supported in part by the University of Pittsburgh, School of Computing and Information, Department of Computer Science, Pitt Cyber, Pitt Momentum fund, PQI Community Collaboration Awards, John C. Mascaro Faculty Scholar in Sustainability, Switzerland NSF award 2000-1-243053, NSF award 2535915, 2610010, DOE Genesis Program, ONR N00014-27-1-2053, Thinking Machines Lab and Cisco Research. This research used resources of the Oak Ridge Leadership Computing Facility, which is a DOE Office of Science User Facility supported under Contract DE-AC05-00OR22725.

\section*{AI Use Disclosure}
AI tools assisted with simulation development, debugging, and manuscript drafting and revision. These uses were directed by the authors. The authors retain full responsibility for the scientific claims, reported results, and final manuscript.

\bibliographystyle{apsrev4-2}
\bibliography{references}

\clearpage
\onecolumngrid
\appendix
\input{appendix_apparatus}

\input{appendix_agent_methods}

\end{document}

%% file: appendix_apparatus.tex
\raggedbottom
\setlength{\intextsep}{6pt}
\section{Detailed Results for Experiment I}
\label{app:apparatus}
\renewcommand{\thesubsection}{\thesection.\arabic{subsection}}

\subsection{Available Apparatus and Control Modes}
\label{app:apparatus-controls}

Table~\ref{tab:available-apparatus} lists the equipment available for Experiment I; not every component is used in each measurement method. Manual denotes robotic manipulation, Electronic denotes controller commands, and Both indicates that both modes apply.

\noindent\begin{minipage}{\textwidth}
\makeatletter\def\@captype{table}\makeatother
\caption{Available optical components and their control modes in Experiment I.}
\label{tab:available-apparatus}
\small
\renewcommand{\arraystretch}{1.0}
\begin{ruledtabular}
\begin{tabular}{p{0.25\textwidth}p{0.51\textwidth}p{0.14\textwidth}}
Component & Available variants or function & Control mode \\
\hline
Scene-observation cameras & One overhead and two oblique cameras ($1600\times1200$ pixels); one wrist camera ($1280\times960$ pixels). Fixed views also provide $640\times480$ depth images. & Both \\
Robotic arm and gripper & Seven-degree-of-freedom UFACTORY xArm7 MuJoCo model from MuJoCo Menagerie & Electronic \\
Light sources & Photon-pair source and coherent alignment lasers & Electronic \\
Mirrors & Protected-silver and dual-band dielectric mirrors; manual and motorized-angle variants & Both \\
Piezo-mounted mirror & Dual-band mirror for fine optical-path scans & Both \\
Beam splitters & Broadband non-polarizing plate and near-infrared non-polarizing cube & Manual \\
Dichroic mirror & Long-pass mirror for separating spectral bands & Manual \\
Wave plates & Half-wave plates for the two spectral bands and an achromatic half-wave plate & Manual \\
Fiber couplers & Band-specific refractive couplers and a broadband reflective coupler; a motorized variant for the longer-wavelength band & Both \\
Sample translation stage & Motorized stage carrying the partially coated sample & Both \\
Optical delay stage & Motorized translation stage with a retroreflector & Both \\
Photon-counting detectors & Fixed superconducting nanowire detectors and movable silicon avalanche-detector modules & Both \\
Time tagger & Detection-time recording and coincidence readout & Electronic \\
Power meters & Free-space detector head and a fixed fiber-input power meter & Both \\
Beam-profiling camera & Beam-intensity imaging for alignment & Both \\
Adjustable iris & Aperture for beam-axis definition and alignment & Manual \\
Beam dump & Passive absorption of unused light & Manual \\
\end{tabular}
\end{ruledtabular}
\end{minipage}

\enlargethispage{\baselineskip}
\vspace{-12pt}
\subsection{Views of the Virtual Laboratory}
\label{app:lab-views}

Fig.~\ref{fig:lab-views} shows rendered views of the virtual laboratory used in the thin-film thickness measurement experiment.

\noindent\begin{minipage}{\textwidth}
\makeatletter\def\@captype{figure}\makeatother
\centering
\begin{minipage}[t]{0.39\textwidth}
\centering
\includegraphics[width=\linewidth]{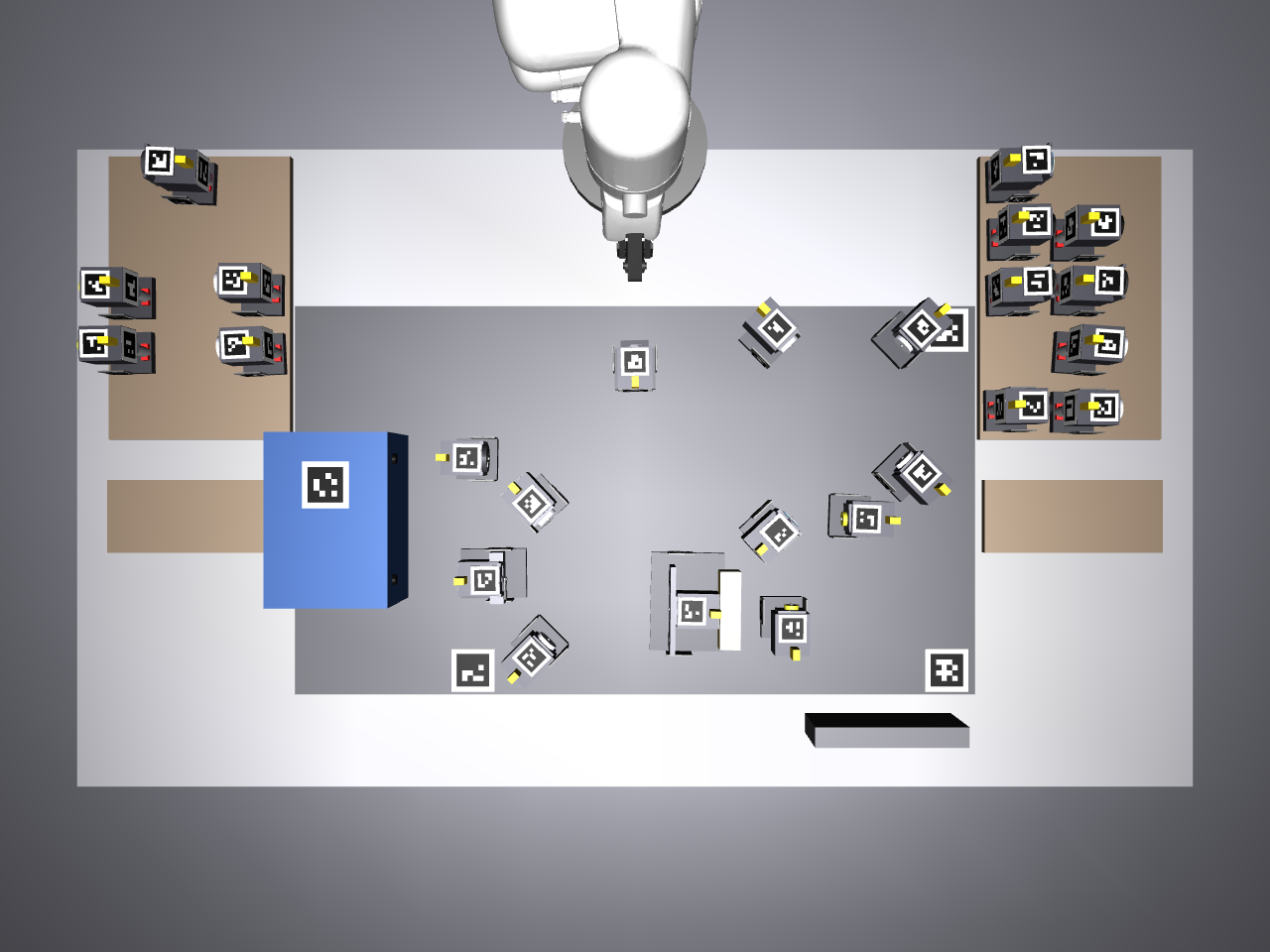}\\[2pt]
(a) Overhead view
\end{minipage}\hfill
\begin{minipage}[t]{0.39\textwidth}
\centering
\includegraphics[width=\linewidth]{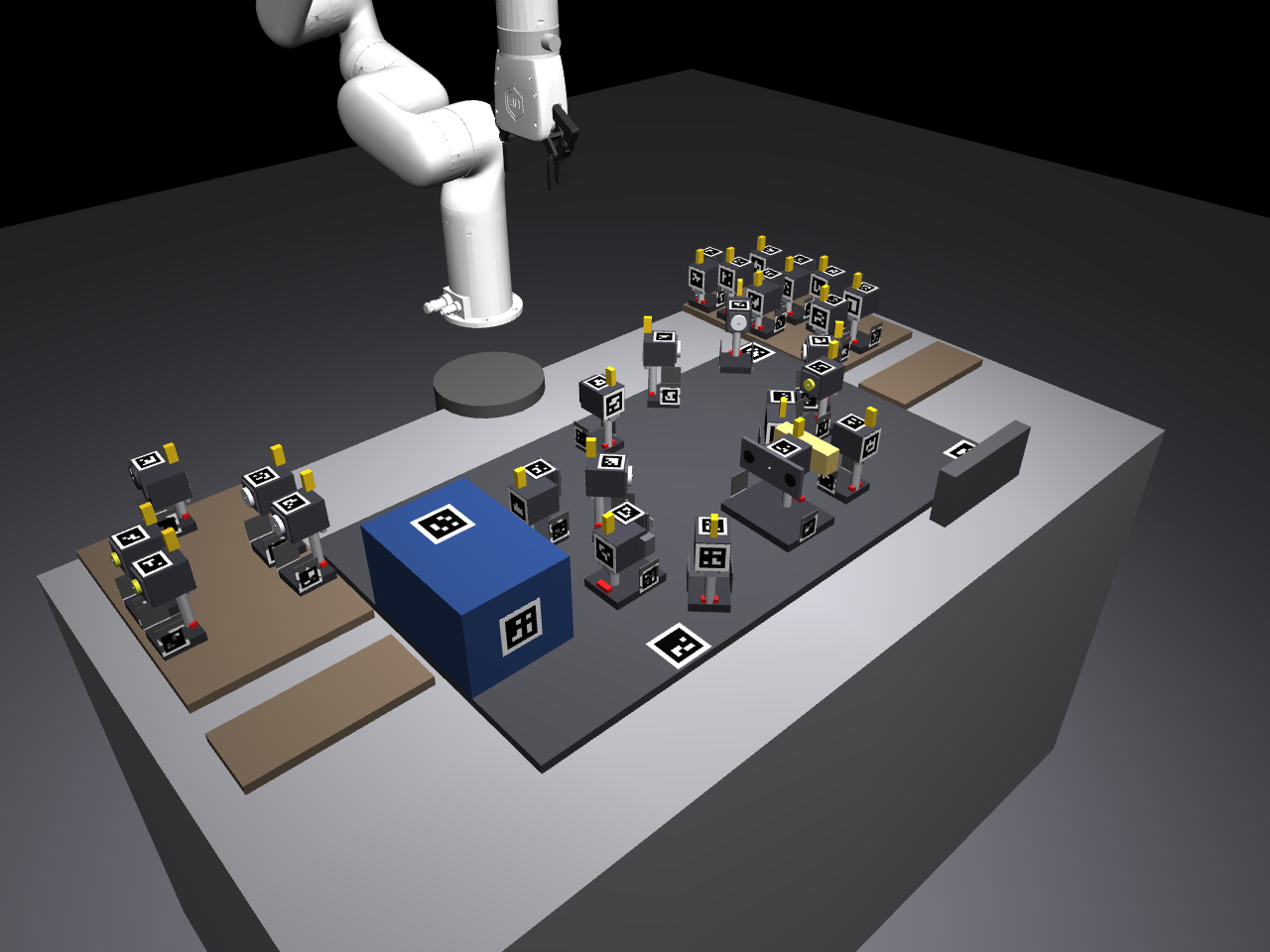}\\[2pt]
(b) Oblique view
\end{minipage}
\caption{Rendered overhead and oblique views of the virtual laboratory used in Experiment I.}
\label{fig:lab-views}
\end{minipage}

\subsection{Measurement Principle of Two-Photon Interferometry}
\label{app:two-photon-principle}

The two-photon measurement follows the principle described by Lualdi et al.~\cite{Lualdi2025}. Energy entanglement produces fringes sensitive to small optical-path changes. Under frequency-independent arm loss, the two interfering alternatives experience the same attenuation, preserving the fringe visibility conditioned on both photons surviving. Coincidence detection separately suppresses uncorrelated background.

A photon pair contains two different central frequencies, corresponding approximately to wavelengths of 810 and 1550~nm. The ideal state entering the two interferometer arms is
\begin{equation}
|\psi\rangle=\frac{1}{\sqrt{2}}\left(|\omega_1\rangle_a|\omega_2\rangle_b+|\omega_2\rangle_a|\omega_1\rangle_b\right).
\end{equation}
The two terms coherently exchange which frequency travels through each arm. A relative delay $\tau=(L_a-L_b)/c$ introduces a relative phase $\Delta\omega\tau$, where $\Delta\omega=\omega_1-\omega_2$. When the photons interfere at a balanced beam splitter, the ideal probability that they exit through different ports is
\begin{equation}
P_C(\tau)=\frac{1}{2}\left[1-\cos(\Delta\omega\tau)e^{-2\sigma^2\tau^2}\right],
\label{eq:two-photon-principle}
\end{equation}
with $\sigma$ the spectral-bandwidth parameter in the Gaussian model of the reference paper. The cosine produces fringes within the broad interference envelope. Their optical-path period is $2\pi c/|\Delta\omega|$, so a large frequency separation gives sensitivity to small path changes without requiring broadband photons. A film introduces different transmission phases $\phi_1(d)$ and $\phi_2(d)$ at the two frequencies. With the film in arm $b$, the fringe phase becomes $\Delta\omega\tau+\phi_2(d)-\phi_1(d)$. Comparing coated and uncoated regions therefore provides information about thickness $d$ through the film-induced fringe shift.

Unequal path loss affects classical and two-photon interference differently. For classical interference at a balanced beam splitter, the output intensity has the form
\begin{equation}
I_{\pm}=\frac{I_a+I_b}{2}\pm\sqrt{I_a I_b}\cos\varphi,
\qquad
V_{\mathrm{cl}}=\frac{2\sqrt{I_a I_b}}{I_a+I_b},
\end{equation}
where $I_a$ and $I_b$ are the intensities reaching the beam splitter and perfect coherence is assumed. Attenuating one initially balanced arm reduces the interference contrast relative to the mean intensity. For the entangled pair, however, both alternatives contain one photon in each arm. If arm transmissions $T_a$ and $T_b$ are independent of frequency over the two bands, the surviving two-photon component is
\begin{equation}
|\psi_{\mathrm{surv}}\rangle_{\mathrm{unnorm}}
=\sqrt{T_aT_b}\,|\psi\rangle.
\end{equation}
After conditioning on both photons surviving, the normalized state and ideal fringe visibility are unchanged. Loss still reduces the pair count and increases statistical uncertainty at fixed acquisition time. Frequency-dependent attenuation or other imperfections can unbalance the two alternatives and reduce visibility, so this robustness is not absolute immunity to loss.

Background suppression comes from the joint detection of the two photons. Wavelength separation at the two output ports gives four detector channels, one for each port--wavelength combination. Pairing detections from different wavelength bands yields four coincidence-count channels: two for photons exiting opposite ports and two for photons exiting the same port. We denote the summed opposite-port counts by $N_C$ and the summed same-port counts by $N_A$, giving the normalized signal $\widehat{P}_C=N_C/(N_C+N_A)$. Both categories count paired detections; $N_A$ does not denote accidental coincidences. A genuine pair produces temporally correlated detections; independent background photons usually do not. For independent stationary event streams with rates $S_\alpha$ and $S_\beta$, the accidental-coincidence rate in a full window width $W$ is approximately
\begin{equation}
R_{\mathrm{acc}}\simeq S_\alpha S_\beta W,
\end{equation}
in the low-occupancy regime, neglecting detector dead time. A narrow window therefore suppresses uncorrelated events while retaining most genuine pairs if it accommodates their timing uncertainty. In contrast, background directly adds to classical intensity or single-count measurements and lowers their fringe contrast. Accidental coincidences, correlated background, and detector saturation can still degrade the two-photon signal. Thus, coincidence readout provides background rejection, while energy entanglement supplies the rapidly varying fringes needed for sensitive optical-delay measurements.

%% file: appendix_agent_methods.tex
\subsection{Measurement Methods in Experiment I}
\label{app:agent-methods}

The tables summarize the submitted method in each run and epoch. Epoch 1 lists the initial agent-developed method; later entries describe changes relative to the preceding epoch in the same run.

\Needspace{16\baselineskip}
\subsubsection{Classical interference}
\label{app:agent-classical}
\begingroup
\renewcommand{\arraystretch}{1.18}
\setlength{\LTcapwidth}{\textwidth}
\setlength{\LTpre}{0.4\baselineskip}
\setlength{\LTpost}{0.6\baselineskip}
\begin{longtable}{|c|@{}c@{}|}
\caption{Changes to the measurement methods under classical interference.}\label{tab:agent-classical}\\
\hline
\makebox[0.065\textwidth]{Run} & \begin{tabular}[c]{c|c}
\makebox[0.07\textwidth]{Epoch} & \parbox[c]{\dimexpr0.865\textwidth-6\tabcolsep-4\arrayrulewidth\relax}{\strut Method or change from the previous epoch\strut} \\
\end{tabular} \\
\hline
\endfirsthead
\multicolumn{2}{c}{\tablename\ \thetable\ (continued)} \\
\hline
\makebox[0.065\textwidth]{Run} & \begin{tabular}[c]{c|c}
\makebox[0.07\textwidth]{Epoch} & \parbox[c]{\dimexpr0.865\textwidth-6\tabcolsep-4\arrayrulewidth\relax}{\strut Method or change from the previous epoch\strut} \\
\end{tabular} \\
\hline
\endhead
\endfoot
\endlastfoot
1 & \begin{tabular}[c]{c|c}
\makebox[0.07\textwidth]{1} & \parbox[c]{\dimexpr0.865\textwidth-6\tabcolsep-4\arrayrulewidth\relax}{\strut Scan the film edge and jointly fit fringe amplitudes and phases.\strut} \\
\hline
\makebox[0.07\textwidth]{2} & \parbox[c]{\dimexpr0.865\textwidth-6\tabcolsep-4\arrayrulewidth\relax}{\strut Switch to three-position amplitude ratios and a thin-film multiple-reflection model.\strut} \\
\hline
\makebox[0.07\textwidth]{3} & \parbox[c]{\dimexpr0.865\textwidth-6\tabcolsep-4\arrayrulewidth\relax}{\strut Repeat uncoated--edge--coated--edge measurements; jointly fit fringes and amplitudes and update calibration.\strut} \\
\hline
\makebox[0.07\textwidth]{4} & \parbox[c]{\dimexpr0.865\textwidth-6\tabcolsep-4\arrayrulewidth\relax}{\strut Increase exposure and improve fit initialization and fallback handling; retain the measurement positions.\strut} \\
\hline
\makebox[0.07\textwidth]{5} & \parbox[c]{\dimexpr0.865\textwidth-6\tabcolsep-4\arrayrulewidth\relax}{\strut Readjust fiber coupling, use short-wavelength channels for the main fit, and revise channel weights.\strut} \\
\hline
\makebox[0.07\textwidth]{6} & \parbox[c]{\dimexpr0.865\textwidth-6\tabcolsep-4\arrayrulewidth\relax}{\strut Add a partial correction based on the difference between measured phase and amplitude-derived phase.\strut} \\
\end{tabular} \\
\hline
2 & \begin{tabular}[c]{c|c}
\makebox[0.07\textwidth]{1} & \parbox[c]{\dimexpr0.865\textwidth-6\tabcolsep-4\arrayrulewidth\relax}{\strut Alternate measurements across uncoated and coated regions; estimate thickness from the phase step.\strut} \\
\hline
\makebox[0.07\textwidth]{2} & \parbox[c]{\dimexpr0.865\textwidth-6\tabcolsep-4\arrayrulewidth\relax}{\strut Switch to a film-edge amplitude profile and calibrate the thickness conversion using train samples.\strut} \\
\hline
\makebox[0.07\textwidth]{3} & \parbox[c]{\dimexpr0.865\textwidth-6\tabcolsep-4\arrayrulewidth\relax}{\strut Increase piezo sampling, revise channel weights, and update calibration; retain the amplitude-profile method.\strut} \\
\hline
\makebox[0.07\textwidth]{4} & \parbox[c]{\dimexpr0.865\textwidth-6\tabcolsep-4\arrayrulewidth\relax}{\strut Concentrate measurements at the edge, jointly fit shared piezo-position errors, and use thin-film optical inversion.\strut} \\
\hline
\makebox[0.07\textwidth]{5} & \parbox[c]{\dimexpr0.865\textwidth-6\tabcolsep-4\arrayrulewidth\relax}{\strut Add a partial spatial phase-gradient correction and strengthen invalid-data handling.\strut} \\
\hline
\makebox[0.07\textwidth]{6} & \parbox[c]{\dimexpr0.865\textwidth-6\tabcolsep-4\arrayrulewidth\relax}{\strut Retain acquisition and analysis; update calibration offsets and the phase-gradient correction coefficient.\strut} \\
\end{tabular} \\
\hline
3 & \begin{tabular}[c]{c|c}
\makebox[0.07\textwidth]{1} & \parbox[c]{\dimexpr0.865\textwidth-6\tabcolsep-4\arrayrulewidth\relax}{\strut Use interleaved phase measurements in the uncoated and coated regions.\strut} \\
\hline
\makebox[0.07\textwidth]{2} & \parbox[c]{\dimexpr0.865\textwidth-6\tabcolsep-4\arrayrulewidth\relax}{\strut Switch to a spatial complex-fringe model and adjust the output fiber-coupler focus.\strut} \\
\hline
\makebox[0.07\textwidth]{3} & \parbox[c]{\dimexpr0.865\textwidth-6\tabcolsep-4\arrayrulewidth\relax}{\strut Jointly fit the phase profile and latent noise in individual acquisition bins.\strut} \\
\hline
\makebox[0.07\textwidth]{4} & \parbox[c]{\dimexpr0.865\textwidth-6\tabcolsep-4\arrayrulewidth\relax}{\strut Increase sampling and fit separate edge widths for the two wavelength bands with shared temporal drift.\strut} \\
\hline
\makebox[0.07\textwidth]{5} & \parbox[c]{\dimexpr0.865\textwidth-6\tabcolsep-4\arrayrulewidth\relax}{\strut Constrain edge widths using train data, revise wavelength-channel weights, and adapt fitting resolution.\strut} \\
\hline
\makebox[0.07\textwidth]{6} & \parbox[c]{\dimexpr0.865\textwidth-6\tabcolsep-4\arrayrulewidth\relax}{\strut Retain the epoch-5 measurement method; tested alternatives are not adopted.\strut} \\
\end{tabular} \\
\hline
\end{longtable}
\endgroup

\Needspace{16\baselineskip}
\subsubsection{High loss}
\label{app:agent-loss}
\begingroup
\renewcommand{\arraystretch}{1.18}
\setlength{\LTcapwidth}{\textwidth}
\setlength{\LTpre}{0.4\baselineskip}
\setlength{\LTpost}{0.6\baselineskip}
\begin{longtable}{|c|@{}c@{}|}
\caption{Changes to the measurement methods under high loss.}\label{tab:agent-loss}\\
\hline
\makebox[0.065\textwidth]{Run} & \begin{tabular}[c]{c|c}
\makebox[0.07\textwidth]{Epoch} & \parbox[c]{\dimexpr0.865\textwidth-6\tabcolsep-4\arrayrulewidth\relax}{\strut Method or change from the previous epoch\strut} \\
\end{tabular} \\
\hline
\endfirsthead
\multicolumn{2}{c}{\tablename\ \thetable\ (continued)} \\
\hline
\makebox[0.065\textwidth]{Run} & \begin{tabular}[c]{c|c}
\makebox[0.07\textwidth]{Epoch} & \parbox[c]{\dimexpr0.865\textwidth-6\tabcolsep-4\arrayrulewidth\relax}{\strut Method or change from the previous epoch\strut} \\
\end{tabular} \\
\hline
\endhead
\endfoot
\endlastfoot
1 & \begin{tabular}[c]{c|c}
\makebox[0.07\textwidth]{1} & \parbox[c]{\dimexpr0.865\textwidth-6\tabcolsep-4\arrayrulewidth\relax}{\strut Use alternating uncoated/coated measurements, a Poisson fit to the four coincidence-count channels, and calibrated optical inversion.\strut} \\
\hline
\makebox[0.07\textwidth]{2} & \parbox[c]{\dimexpr0.865\textwidth-6\tabcolsep-4\arrayrulewidth\relax}{\strut Readjust optical alignment to increase coincidence counts; use denser spatial sampling and a linear-drift fit.\strut} \\
\hline
\makebox[0.07\textwidth]{3} & \parbox[c]{\dimexpr0.865\textwidth-6\tabcolsep-4\arrayrulewidth\relax}{\strut Add a sample position and use a robust weighted fit.\strut} \\
\hline
\makebox[0.07\textwidth]{4} & \parbox[c]{\dimexpr0.865\textwidth-6\tabcolsep-4\arrayrulewidth\relax}{\strut Retain acquisition; update the calibration offset and restore settings after an alignment probe.\strut} \\
\hline
\makebox[0.07\textwidth]{5} & \parbox[c]{\dimexpr0.865\textwidth-6\tabcolsep-4\arrayrulewidth\relax}{\strut Retain acquisition; add weak shrinkage toward the center of the thickness range and robust analysis.\strut} \\
\hline
\makebox[0.07\textwidth]{6} & \parbox[c]{\dimexpr0.865\textwidth-6\tabcolsep-4\arrayrulewidth\relax}{\strut Adjust the half-wave plate and update the calibration offset; retain the remaining method.\strut} \\
\end{tabular} \\
\hline
2 & \begin{tabular}[c]{c|c}
\makebox[0.07\textwidth]{1} & \parbox[c]{\dimexpr0.865\textwidth-6\tabcolsep-4\arrayrulewidth\relax}{\strut Estimate the phase step from four coincidence-count channels and invert the thin-film response.\strut} \\
\hline
\makebox[0.07\textwidth]{2} & \parbox[c]{\dimexpr0.865\textwidth-6\tabcolsep-4\arrayrulewidth\relax}{\strut Improve fiber alignment; use three sensitive phase settings and closer uncoated/coated positions.\strut} \\
\hline
\makebox[0.07\textwidth]{3} & \parbox[c]{\dimexpr0.865\textwidth-6\tabcolsep-4\arrayrulewidth\relax}{\strut Shorten measurement visits, update calibration, and add detector resets.\strut} \\
\hline
\makebox[0.07\textwidth]{4} & \parbox[c]{\dimexpr0.865\textwidth-6\tabcolsep-4\arrayrulewidth\relax}{\strut Fix the phase working points, shorten acquisition bins, and average two pairs of sample positions.\strut} \\
\hline
\makebox[0.07\textwidth]{5} & \parbox[c]{\dimexpr0.865\textwidth-6\tabcolsep-4\arrayrulewidth\relax}{\strut Sample four positions in each region twice; add an offset and thickness-dependent calibration.\strut} \\
\hline
\makebox[0.07\textwidth]{6} & \parbox[c]{\dimexpr0.865\textwidth-6\tabcolsep-4\arrayrulewidth\relax}{\strut Retain acquisition; update only the calibration intercept using accumulated train data.\strut} \\
\end{tabular} \\
\hline
3 & \begin{tabular}[c]{c|c}
\makebox[0.07\textwidth]{1} & \parbox[c]{\dimexpr0.865\textwidth-6\tabcolsep-4\arrayrulewidth\relax}{\strut Align output coupling and use alternating sample regions, phase dithering, and a joint Poisson fit.\strut} \\
\hline
\makebox[0.07\textwidth]{2} & \parbox[c]{\dimexpr0.865\textwidth-6\tabcolsep-4\arrayrulewidth\relax}{\strut Move measurements farther from the film edge to reduce mixed-region bias.\strut} \\
\hline
\makebox[0.07\textwidth]{3} & \parbox[c]{\dimexpr0.865\textwidth-6\tabcolsep-4\arrayrulewidth\relax}{\strut Optimize alignment using coincidence counts rather than single-detector counts; retain the sampling geometry.\strut} \\
\hline
\makebox[0.07\textwidth]{4} & \parbox[c]{\dimexpr0.865\textwidth-6\tabcolsep-4\arrayrulewidth\relax}{\strut Return to nearer-edge positions and add a correction based on the fitted fringe frequency.\strut} \\
\hline
\makebox[0.07\textwidth]{5} & \parbox[c]{\dimexpr0.865\textwidth-6\tabcolsep-4\arrayrulewidth\relax}{\strut Remove the frequency-based correction after new train measurements fail to support it; update calibration.\strut} \\
\hline
\makebox[0.07\textwidth]{6} & \parbox[c]{\dimexpr0.865\textwidth-6\tabcolsep-4\arrayrulewidth\relax}{\strut Use two matched phase working points and a fixed fringe frequency; add a small thickness-dependent calibration.\strut} \\
\end{tabular} \\
\hline
\end{longtable}
\endgroup

\Needspace{16\baselineskip}
\subsubsection{Strong background}
\label{app:agent-background}
\begingroup
\renewcommand{\arraystretch}{1.18}
\setlength{\LTcapwidth}{\textwidth}
\setlength{\LTpre}{0.4\baselineskip}
\setlength{\LTpost}{0.6\baselineskip}
\begin{longtable}{|c|@{}c@{}|}
\caption{Changes to the measurement methods under strong background.}\label{tab:agent-background}\\
\hline
\makebox[0.065\textwidth]{Run} & \begin{tabular}[c]{c|c}
\makebox[0.07\textwidth]{Epoch} & \parbox[c]{\dimexpr0.865\textwidth-6\tabcolsep-4\arrayrulewidth\relax}{\strut Method or change from the previous epoch\strut} \\
\end{tabular} \\
\hline
\endfirsthead
\multicolumn{2}{c}{\tablename\ \thetable\ (continued)} \\
\hline
\makebox[0.065\textwidth]{Run} & \begin{tabular}[c]{c|c}
\makebox[0.07\textwidth]{Epoch} & \parbox[c]{\dimexpr0.865\textwidth-6\tabcolsep-4\arrayrulewidth\relax}{\strut Method or change from the previous epoch\strut} \\
\end{tabular} \\
\hline
\endhead
\endfoot
\endlastfoot
1 & \begin{tabular}[c]{c|c}
\makebox[0.07\textwidth]{1} & \parbox[c]{\dimexpr0.865\textwidth-6\tabcolsep-4\arrayrulewidth\relax}{\strut Alternate uncoated/coated visits; scan the piezo and fit cubic temporal drift with a calibrated offset.\strut} \\
\hline
\makebox[0.07\textwidth]{2} & \parbox[c]{\dimexpr0.865\textwidth-6\tabcolsep-4\arrayrulewidth\relax}{\strut Reduce region separation, use an uncoated--coated--coated--uncoated order, and update calibration.\strut} \\
\hline
\makebox[0.07\textwidth]{3} & \parbox[c]{\dimexpr0.865\textwidth-6\tabcolsep-4\arrayrulewidth\relax}{\strut Improve output coupling, split phase settings into shorter bins, and add a linear calibration correction.\strut} \\
\hline
\makebox[0.07\textwidth]{4} & \parbox[c]{\dimexpr0.865\textwidth-6\tabcolsep-4\arrayrulewidth\relax}{\strut Adjust optical settings, reduce the number of phase settings, and increase short-bin sampling and exposure.\strut} \\
\hline
\makebox[0.07\textwidth]{5} & \parbox[c]{\dimexpr0.865\textwidth-6\tabcolsep-4\arrayrulewidth\relax}{\strut Further reduce region separation and introduce linear and quadratic calibration terms.\strut} \\
\hline
\makebox[0.07\textwidth]{6} & \parbox[c]{\dimexpr0.865\textwidth-6\tabcolsep-4\arrayrulewidth\relax}{\strut Retain acquisition and apparatus settings; replace least squares with Poisson maximum likelihood and recalibrate.\strut} \\
\end{tabular} \\
\hline
2 & \begin{tabular}[c]{c|c}
\makebox[0.07\textwidth]{1} & \parbox[c]{\dimexpr0.865\textwidth-6\tabcolsep-4\arrayrulewidth\relax}{\strut Use a multi-position round-trip scan, phase maximum likelihood, quadratic drift, and linear calibration.\strut} \\
\hline
\makebox[0.07\textwidth]{2} & \parbox[c]{\dimexpr0.865\textwidth-6\tabcolsep-4\arrayrulewidth\relax}{\strut Fix the piezo setting, shorten acquisition bins, reduce spatial positions, and combine phase-step and extrapolation estimates.\strut} \\
\hline
\makebox[0.07\textwidth]{3} & \parbox[c]{\dimexpr0.865\textwidth-6\tabcolsep-4\arrayrulewidth\relax}{\strut Use three phase working points and shorter bins; increase scan repetitions and the extrapolation weight.\strut} \\
\hline
\makebox[0.07\textwidth]{4} & \parbox[c]{\dimexpr0.865\textwidth-6\tabcolsep-4\arrayrulewidth\relax}{\strut Improve fiber alignment and focus; select phase working points during measurement.\strut} \\
\hline
\makebox[0.07\textwidth]{5} & \parbox[c]{\dimexpr0.865\textwidth-6\tabcolsep-4\arrayrulewidth\relax}{\strut Reduce sampling to three positions and use linear extrapolation; a disk failure limits development evidence.\strut} \\
\hline
\makebox[0.07\textwidth]{6} & \parbox[c]{\dimexpr0.865\textwidth-6\tabcolsep-4\arrayrulewidth\relax}{\strut Retain acquisition; reduce data-read page size to avoid output failures and update calibration.\strut} \\
\end{tabular} \\
\hline
3 & \begin{tabular}[c]{c|c}
\makebox[0.07\textwidth]{1} & \parbox[c]{\dimexpr0.865\textwidth-6\tabcolsep-4\arrayrulewidth\relax}{\strut Use the film-edge amplitude dip and phase localization instead of a distant phase-step measurement.\strut} \\
\hline
\makebox[0.07\textwidth]{2} & \parbox[c]{\dimexpr0.865\textwidth-6\tabcolsep-4\arrayrulewidth\relax}{\strut Use four quadrature phase settings and longer visits; combine amplitude-ratio and background-corrected mean-count estimates.\strut} \\
\hline
\makebox[0.07\textwidth]{3} & \parbox[c]{\dimexpr0.865\textwidth-6\tabcolsep-4\arrayrulewidth\relax}{\strut Readjust output coupling to increase counts; retain the sampling geometry and recalibrate.\strut} \\
\hline
\makebox[0.07\textwidth]{4} & \parbox[c]{\dimexpr0.865\textwidth-6\tabcolsep-4\arrayrulewidth\relax}{\strut Correct the coated-region factor in the mean-count estimator and revise amplitude normalization.\strut} \\
\hline
\makebox[0.07\textwidth]{5} & \parbox[c]{\dimexpr0.865\textwidth-6\tabcolsep-4\arrayrulewidth\relax}{\strut Optimize fiber focus; interleave edge and reference visits and interpolate reference drift using a random-walk model.\strut} \\
\hline
\makebox[0.07\textwidth]{6} & \parbox[c]{\dimexpr0.865\textwidth-6\tabcolsep-4\arrayrulewidth\relax}{\strut Retain acquisition and apparatus settings; update estimator weights and calibration using accumulated train data.\strut} \\
\end{tabular} \\
\hline
\end{longtable}
\endgroup